\documentclass[11pt,a4paper]{article}

\usepackage{amsfonts,amsmath,amssymb}
\usepackage{mathrsfs}
\usepackage{amscd}
\usepackage{color}
\usepackage{graphicx}

\font\tengot=eufm10  \font\sevengot=eufm7  \font\fivegot=eufm5

\newfam\gotfam
\textfont\gotfam=\tengot
\scriptfont\gotfam=\sevengot
\scriptscriptfont\gotfam=\fivegot

\begin{document}
	
\title {Axially symmetric ghost stars}

\author{L. Herrera\thanks{Instituto Universitario de F\'\i sica Fundamental y Matem\'aticas (IUFFyM), Universidad de Salamanca, Salamanca 37007, Spain. E-mail address: lherrera@usal.es} , \   J.L. Hern\'andez--Pastora\thanks{Departamento de Matem\'atica Aplicada, Facultad de Ciencias, Universidad de Salamanca,  and  IUFFyM. https://ror.org/02f40zc51, Universidad de Salamanca, Salamanca 37007, Spain.   E-mail address: jlhp@usal.es. ORCID:orcig.org/0000-0002-3958-6083} , J. Ospino \thanks{Departamento de Matem\'atica Aplicada, Facultad de Ciencias, Universidad de Salamanca, Salamanca 37007, Spain and  IUFFyM.  E-mail address: j.ospino@usal.es.} and A. Di Prisco \thanks{Escuela de Fis\'ica, Facultad de Ciencias. Universidad Central de Venezuela, Caracas 1050, Venezuela.   E-mail address: alicia.diprisco@ucv.ve.} \\}

\date{\today}

\maketitle

\vspace{-10mm}

\begin{abstract}
We present static axially symmetric  fluid distributions not producing gravitational field outside their boundaries (i.e. fluid sources  which match smoothly on the boundary surface to Minkowski space--time).  These solutions provide further examples of ghost stars.   A specific model is fully described, and its physical and geometrical properties  are analyzed in detail. This includes the multipole moment structure of the source and its complexity factors, both of which vanish for our solution.
 \end{abstract}

PACS numbers:  04.20.Cv, 04.20.-q, 4.20.Ha, 95.30.Sf.

\large


\section{Introduction}
In a recent paper \cite{1}, a method proposed in \cite{weylsol} to obtain interior solutions representing axially symmetric fluids  was used to obtain axially symmetric fluid distributions which match smoothly to  the Schwarzschild line element. In this work we want to specialize this approach to find static axially symmetric fluid distributions matching smoothly to Minkowski space-time. Examples of spherically symmetric fluid distributions not producing gravitational field outside their boundaries have been presented in \cite{3n,1n,1nb}  (see also \cite{na} for more recent developments), motivated by early ideas of Zeldovich \cite{2n,2nb}, and were called ``ghost stars''  in \cite{1n} due to  their similarity with some Einstein--Dirac  neutrinos (named  ghost neutrinos)  which do not produce gravitational field but  still are characterized by  non vanishing current density \cite{gn,gn1,gn2}.

In this work we extend the concept of ``ghost stars''  to the axially symmetric static case. We recall that all vacuum axially symmetric static space-times are
encompassed within the family  of the Weyl exterior solutions \cite{weyl1}-\cite{weyln}.  These solutions also arise  when one considers the matter configurations with flat or ``slab'' symmetry (see, \cite{refref} and references therein).

The rationale behind the idea to consider axially symmetric self-gravitating objects is based on the well known fact  that the only regular static and asymptotically flat vacuum spacetime possessing a regular horizon is the Schwarzschild solution \cite{israel}, while all the others Weyl exterior solutions  exhibit singularities in the  curvature invariants (as the boundary of the source approaches the horizon). This in turn implies that, for  very compact objects, a bifurcation appears between any finite perturbation of Schwarzschild spacetime and any Weyl solution, even when the latter is characterized by parameters arbitrarily close to those corresponding to spherical symmetry (see \cite{i2}-\cite{i5} and references therein for a discussion on this point).

Thus, even if we know that observational evidence seems to suggest  that deviations from spherical symmetry in compact
self-gravitating objects are
likely to be incidental rather than basic features of these systems, it should  be clear that a rigorous description of static axially symmetric sources, including finding exact analytical solutions, is a praiseworthy endeavor.

The jump of symmetry across the boundary surface observed in the examples studied  in \cite{1} as well as in the solutions presented in this work, is a known situation in GR. Let us just mention that  the Szekeres  space--time \cite{1c,2},  representing   dust models and which have no Killing vectors (not even a time--like one) \cite{3,5},  may be matched smoothly to  the Schwarzschild line element \cite{3}.
 
It is the purpose of this work, to find a family of solutions representing non--spherical, static,  fluid distributions which match smoothly to the Minkowski line element. 
The solutions are obtained by using the general approach outlined in \cite{weylsol}, and may be regarded as  special cases of the solutions proposed in \cite {1}.

The evolution of the research work leading to  the method to obtain the solutions we are looking for here, may be briefly  summarized as follows.  

First, in \cite{epjc_vol}, both  anisotropic and isotropic spherical interior sources, as well as non-spherical ones, smoothly matched  to  any exterior Weyl solution were obtained. The metric functions of the interior solution for a global model could be integrated in terms of some combination of the energy--momentum tensor components, thereby providing some constraints  
on the source,  derived from the exterior gravitational field of the global metric. Thus, these constraints
can be expressed in terms of the gravitational field
which is matching the interior solution.

Next, the method
proposed in \cite{weylsol} allows us to construct a well matched interior
metric for any exterior solution of the Weyl family,  allowing in turn  to  find out,  for any vacuum solution, how the energy--momentum tensor of axially symmetric static sources is
affected by different physical characteristics of the gravitational field outside the source.

These results  were applied in \cite{1} to find   axially symmetric interior solutions smoothly matched to   the Schwarzschild line element.

Finally in this work the procedure described in \cite{1} is specialized to the case when the exterior space-time is described by the Minkowski line element.

As in \cite{1}, the physical and geometric properties of the obtained source are analyzed in detail. The electric Riemann tensor is calculated. It is shown that it vanishes, implying that the source is characterized by the  vanishing of its complexity factors.

The  link between the relativistic multipole moments ($RMM$) \cite{geroch} and   the matter content of the source has been established in 
 \cite{whom}, through explicit expressions of the  $RMM$ in terms of integrals over the space--time filled by the source. Thus, for any interior solution smoothly matched to any vacuum metric, it is possible to calculate the corresponding $RMM$, which of course correspond to the exterior vacuum solution. As expected, we prove that our interior solution  smoothly matched  to the Minkowski space--time,  is characterized by the vanishing of all multipole moments.

\section{The interior metric and its sources}

Let us  first  briefly summarize  the general method developed in \cite{weylsol}  to find matchable solutions to the Weyl space--time, and describe the general conventions and notation.
 
\subsection{The global model}

In  \cite{weylsol} a global metric is obtained for any static solution belonging to the Weyl family of the axisymmetric vacuum Einstein equations. The exterior line element written in Erez-Rosen coordinates is given by 
\begin{eqnarray}
ds^2_E&=&-e^{2\psi(r,y)}dt^2+e^{-2\psi+2\left[\Gamma(r,y)-\Gamma^s
	\right]}dr^2+
e^{-2\left[\psi-\psi^s\right]+
	2\left[\Gamma(r,y)-\Gamma^s\right]}r^2 d\theta^2\nonumber \\
&+& e^{-2\left[\psi-\psi^s\right]} r^2\sin^2\theta d\phi^2,
\label{exterior}
\end{eqnarray}
where $y\equiv \sin \theta$ and  $\psi^s$ and $\Gamma^s$ are the metric functions corresponding to the Schwarzschild solution, namely,
\begin{equation}
\psi^s=\frac 12 \ln \left(\frac{r-2\sigma}{r}\right) \, \quad \Gamma^s=-\frac 12 \ln \left[\frac{(r-\sigma)^2-y^2\sigma^2}{r(r-2\sigma)}\right],
\label{schwfun}
\end{equation}
 the parameter $\sigma$ being easily identified as the Schwarzschild mass.

For the interior axially symmetric line element we shall assume

\begin{equation}
ds^2_I=-e^{2 \hat a} Z(r)^2 dt^2+\frac{e^{2\hat g-2\hat a}}{A(r)} dr^2+e^{2\hat g-2\hat a}r^2 d\theta^2+e^{-2 \hat a}r^2 \sin^2\theta d\phi^2,
\label{interior}
\end{equation}
where $\hat a(r,\theta)$ and  $\hat g(r,\theta)$ are functions whose general form is given below in (\ref{aygsimple}). Also,  $A(r)\equiv 1-pr^2$  and  $Z\equiv\displaystyle{ \frac 32 \sqrt{A(r_{\Sigma})}-\frac 12 \sqrt{ A(r)}}$, where $p$ is an arbitrary constant and the boundary surface of the source is defined by $r=r_{\Sigma}=const.$ 

The choice of  both functions ($A$ and $Z$) is suggested by the condition  that they correspond to the well known incompressible (homogeneous energy density) perfect fluid sphere ($\hat g=\hat a=0$).  Besides they are  restricted by the matching conditions requiring that $A(r_ {\Sigma})=1-\frac{2\sigma}{r_ {\Sigma}}$, $Z_{\Sigma}=\sqrt{A_{\Sigma}}$ and $Z^{\prime}_{\Sigma}=\frac{\sigma}{r_{\Sigma}^2\sqrt{A_{\Sigma}}}$ which lead to  $p=\displaystyle{\frac{2\sigma}{r_{\Sigma}^3}}$. 

The other interior metric functions must hold the following form:
\begin{eqnarray}
\hat a(r,\theta)&=&\hat \psi_{\Sigma} s^2(3-2s)   +r_{\Sigma}\hat \psi^{\prime}_{\Sigma}s^2(s-1)+(r-r_{\Sigma})^2F(r,\theta),\nonumber \\
\hat g(r,\theta)&=&\hat \Gamma_{\Sigma} s^3(4-3s)   +r_{\Sigma}\hat \Gamma^{\prime}_{\Sigma}s^3(s-1)+(r-r_{\Sigma})^2G(r,\theta).
\label{aygsimple}
\end{eqnarray}
with $s\equiv r/r_{\Sigma} \in \left[0,1\right]$, $\hat \psi \equiv \psi-\psi^s$, $\hat \Gamma \equiv \Gamma-\Gamma^s$ and  $F(r,\theta)$, $G(r, \theta)$ arbitrary functions satisfying  the following constraints: $F(0,\theta)=F^{\prime}(0,\theta)=0$,  $G(0,\theta)=G^{\prime}(0,\theta)=G^{\prime \prime}(0,\theta)=0$ derived from the matching conditions and regular behavior at the origin (prime denotes derivative with respect to $r$, and subindex $\Sigma$ denotes evaluation on the boundary surface).

Next, we shall assume Minkowski space-time ($\psi=\Gamma=0$) outside of the source, and  take the parameter $\sigma=0$. Thus   the exterior line element, corresponding to a flat space-time, reads
\begin{equation}
ds^2_E=-dt^2+dr^2+
r^2 d\theta^2+ r^2\sin^2\theta d\phi^2.
\label{exteriorM}
\end{equation}

For our line element  (\ref{interior}) with $\sigma=0, p=0$ (implying  $A=Z=1$),  the non-vanishing  components of the energy--momentum tensor read (see \cite{weylsol} for details)

\begin{equation}
-T^0_0=\gamma \left(\hat p_{zz}+E\right),
T^1_1=-T^2_2=\gamma \left(-\hat p_{xx}\right),
T^3_3=\gamma \left(-\hat p_{zz}\right),
\label{presionPpzero}
\end{equation} and \begin{equation}
T_1^2=g^{\theta\theta}T_{12}=-\frac{\gamma}{r^2}\left[2 {\hat a}_{,\theta}  \hat a^{\prime}-\hat g^{\prime}\frac{\cos\theta}{\sin\theta}-\frac{\hat g_{,\theta}}{r}\right],
\label{pxypzero}
\end{equation} with 
\begin{eqnarray}
&&E=2 \left[\hat a^{\prime \prime}+2\frac{\hat a^{\prime}}{r}+\frac{{\hat a}_{,\theta \theta} }{r^2}+\frac{{\hat a}_{,\theta} }{r^2}\frac{\cos \theta}{\sin \theta}\right],\nonumber\\
&&\hat p_{xx}=-\frac{{\hat a}_{,\theta} ^2}{r^2}-\frac{\hat g^{\prime}}{r}+\hat a^{\prime 2}+\frac{{\hat g}_{,\theta} }{r^2}\frac{\cos \theta}{\sin \theta}, \nonumber \\
&&\hat p_{zz}=-\frac{{\hat a}^2_{,\theta} }{r^2}-\frac{\hat g^{\prime}}{r}-\hat a^{\prime 2}-\frac{{\hat g}_{,\theta \theta} }{r^2}-\hat g^{\prime \prime},
\label{eegeneraldetpzero}
\end{eqnarray}
with
\begin{equation}
\hat a(r,\theta)=(r-r_{\Sigma})^2F(r,\theta),\quad 
\hat g(r,\theta)=(r-r_{\Sigma})^2G(r,\theta),
\label{aygsimplepzero}
\end{equation}
where $\displaystyle{\gamma\equiv \frac{e^{2\hat a-2\hat g}}{8 \pi}}$, subscript denotes derivative with respect to the angular variable $\theta$, and $\hat p_{zz}, \hat p_{xx}, T_1^2, E$ describe deviations from the vacuum flat space-time.
Indeed, if $\hat a=\hat g=0$,  then $E=\hat p_{xx}=\hat p_{zz}=0$ and we recover the vacuum energy-momentum tensor.

We will now address the mass  problem.  As  is well known, both the Komar mass $M_K$ \cite{Ko} and the  Tolman $M_T$ \cite{To} coincide for the static case, and furthermore when matching the source   to  the Schwarzschild line element, the mass evaluated at the boundary surface coincides with the parameter $M$ of the Schwarzschild metric. Accordingly we expect the total mass of the source to vanish when matching to the Minkowski line element. As we shall see below this is in fact what happens.

From their definition, we have 
\begin{equation}
M_T=M_K=-\frac{1}{4\pi}\int_V\sqrt{-\bf{g}}R^0_0 d^3x=\int_V(-T^0_0+\hat T)\sqrt{-\bf{g}} d^3x,
\end{equation}
with an obvious notation. 

In order to compute the above integral we have to take into account the following relationships 
\begin{equation}
\sqrt{-\bf{g}}R^0_0=\partial_k\left(\sqrt{-\bf{g}}g^{00}\Gamma^k_{00}\right)=
-\partial_k \left(-\sqrt{\bf{\hat g}}\hat g^{kj}\partial_j\sqrt{-g_{00}}\right)=-\sqrt{\bf{\hat g}}\hat{\Delta} \sqrt{-g_{00}},
\end{equation}
where  $\bf{\hat{g}}$ is the determinant of the three--space metric and  $\hat{\Delta}$ denots the second kind  Beltrami operator for such a metric. 

Next, we may write for the  total mass
\begin{equation}
M_T=M_K=\int_V \rho_T \hat{\eta},
\end{equation}
where  $\hat{\eta}=\sqrt{\bf{\hat g}}d^3x$ is the three dimensional volume element, and  
\begin{equation}
\hat{\Delta}\sqrt{-g_{00}}=4\pi \rho_T \ , \quad \rho_T\equiv\sqrt{-g_{00}} \left(-T^0_0+\hat T\right).
\end{equation}

Then, applying the  Gauss theorem
\begin{equation}
M_T=\frac{1}{4\pi}\int_V\hat{\Delta}\sqrt{-g_{00}}\hat{\eta}=\frac{1}{4\pi}\int_V\partial_k\left(-\sqrt{\bf{\hat g}}\hat g^{kj}\partial_j\sqrt{-g_{00}}\right) d^3x,
\end{equation}
we obtain for the total mass
\begin{equation}
M_T=\frac{1}{4\pi}\int_{\partial V}\sqrt{\bf{\hat g}} \hat g^{kj}\partial_j\sqrt{-g_{00}} \ n_k d\sigma.
\end{equation}

The above expression yields in the general case, for our interior (\ref{interior}),
\begin{equation}
M_T=\frac 12 \int_0^{\pi} r_{\Sigma}^2 \sqrt{A_{\Sigma}}e^{-\hat{\psi}_{\Sigma}}
\left(\partial_r\sqrt{-g_{00}}\right)_{\Sigma} \sin\theta d \theta .
\label{mn1}
\end{equation}

Finally, in the particular case when we match our source with the  Minkowski exterior, we have  $\hat{\psi}_{\Sigma}=-\psi^s_{\Sigma}$, and  $\sqrt{-g_{00}}=e^{-\psi^s}Z(r)$, implying   
\begin{equation}
\left(\partial_r \sqrt{-g_{00}}\right)_{\Sigma}=e^{-\psi^s}_{\Sigma}\left(
\frac{\sigma}{r_{\Sigma}^2\sqrt{A_{\Sigma}}}-\sqrt{A_{\Sigma}} {\psi}_{\Sigma}^{s \prime}\right),
\end{equation}
then, feeding back the above expression into (\ref{mn1}) we obtain
\begin{equation}
M_T=M_K=\sigma-\frac{1}{2}r_{\Sigma}^2 A_{\Sigma}\int_0^{\pi}d\theta \sin\theta \frac{\sigma}{r_{\Sigma}(r_ {\Sigma}-2\sigma)}=0,
\label{masatotal}
\end{equation}	
since $\sigma=0$.
\section{Sources matched to the Minkowski space--time}

We shall now tackle the problem motivating this work. If we consider that the exterior space--time is  flat (Minkowski), then the interior metric functions (\ref{aygsimple}) would be
\begin{equation}
\hat a(r,\theta)=(r-r_{\Sigma})^2F(r,\theta),\qquad 
\hat g(r,\theta)=(r-r_{\Sigma})^2G(r,\theta),
\label{aygsimpleschw}
\end{equation}
which are  arbitrary,  with the only constraints mentioned above for $F$ and $G$ with respect to their behaviour at the origin of coordinates as well as conditions preserving the good behaviour at the symmetry axis $G(r,\theta=0)=0$.

\vskip 2mm
\subsection{Spherical sources}
In order to illustrate our approach, it could be convenient to  describe briefly the case of spherical sources. In  such a case,  the arbitrary functions $F$ and $G$ only depend on the radial coordinate, and we are able to recover the results obtained in \cite{luisJA}, \cite{Lake} where a procedure for obtaining all the interior solutions for the spherical case has been established. 

As discussed in \cite{1} in this case  the metric function $G(r)$ must be chosen to be null.

Besides, for this case we have that $p_{xx} =- p_{zz}$ and then $T^2_2 = T^3_3$, which means that only two independent main stresses  exist in this case. These are  usually denoted in the literature as $p_r$ (radial pressure) and $p_{\bot}$ (tangential pressure) whenever the above spherical gauge is used for the coordinates. 

The relationship for the metric functions and the  anisotropy $\Pi(r)$ is the following:
\begin{eqnarray}\label{bc}
e^{\nu[\hat r=\hat r(r)]}&=&e^{2\hat a},   \\
e^{-\lambda(\hat r=\hat r(r))}&=&(1-r\hat a^{\prime})^2,  \label{bc1} \\
\Pi(r)&\equiv &8\pi (T_1^1-T_2^2)  \equiv 8\pi(p_r-p_{\bot})=-e^{2\hat a} \hat a^{\prime 2}, 
\label{rela}
\end{eqnarray}
where $\nu$ and $\lambda$ denote the two spherical metric functions appearing in \cite{luisJA} $g_{00}=-e^{\nu}$ and $g_{11}=e^{\lambda}$.

 The total mass of the source may be  calculated by means of the following integration over the volumen interior to the boundary-surface
 \begin{equation}
 M\equiv \int_V T_0^0 d^3x=\int_0^{r_{\Sigma}}dr\int_0^{\pi}d \theta\int_0^{2\pi} d\phi \ r^2 \sin\theta \ e^{2\hat g-2\hat a}\ T_0^0.
 \label{masaToo}
 \end{equation}

In the  spherically  symmetric case, the above integral becomes    \begin{equation}
M=\frac 12\int^{r_{\Sigma}}_0 \left[ r^2\hat a^{\prime 2} -2r^2 \hat a^{\prime \prime}-4 r \hat a^{\prime} \right] dr,
\label{Masazerogzero}
\end{equation}
where  (\ref{presionPpzero}) has been used.

Next, let us notice  that the two negative terms into the above integral are just the derivative of $2r^2 \hat a^{\prime}$ with respect to the radial coordinate, and therefore the integral of these terms leads to evaluate $2r^2 \hat a^{\prime}$  at $r=0$ and  $r=r_{\Sigma}$ where it is null as it follows  from  (\ref{aygsimpleschw}).

Accordingly, the integral above becomes, 
\begin{equation}
M=\frac 12\int^{r_{\Sigma}}_0  r^2\hat a^{\prime 2} dr ,
\label{Masazerogzero2}
\end{equation}
and therefore we conclude that $M$ cannot be zero unless $a^{\prime}=0$. However, this last condition would imply because  (\ref{aygsimpleschw}) and the regularity of $F$ at the boundary surface,  that $a^{\prime}=0$ , which in turn, using (\ref{bc}), (\ref{bc1}) imply that our interior is also  the flat (Minkowski) space-time. 

This result is expected since from the beginning we imposed the condition that our source  describes an incompressible isotropic fluid in the spherically symmetric case. However, non-trivial ghost star requires that the energy density of the fluid distribution be negative in some parts of the distribution, which of course is impossible to satisfy if the energy density is homogeneous

\vskip 2mm
\subsection{Non-spherical sources}
Let us now turn to  the general non-spherical case,  for which  the expressions are more complicated. Thus, in order to specify our model we need  to introduce some simplifying assumptions: We shall assume  $\hat a=0$. Such a choice is justified by the fact that, as can be seen from (\ref{eegeneraldetpzero}), the function $E$ only depends on the metric function $\hat a$ and furthermore it implies $F(r, \theta)=0$.

The expressions for  $p_{xx}$ and $p_{zz}$ for our model  ($E=0$) are
\begin{eqnarray}
\hat p_{xx}&=&\frac{-1}{r_{\Sigma}^2}\left[y\frac{\hat g_{y}}{s^2}+\frac{\hat g_s}{s}\right],\nonumber \\
\hat p_{zz}&=&\frac{1}{r_{\Sigma}^2}\left[ -(1-y^2)\frac{\hat g_{yy}}{s^2}+y\frac{\hat g_{y}}{s^2}-\frac{\hat g_s}{s}-\hat g_{ss}\right],\nonumber \\
\label{laspes}
\end{eqnarray}
where $y\equiv \cos\theta$, $s\equiv r/r_{\Sigma}$, and subscripts denote derivatives.

The total mass $M$ integrated over the volumen of the source from (\ref{masaToo}) and (\ref{presionPpzero})  is now:
\begin{equation}
M=2 \pi \int_0^{r_{\Sigma}}r^2 T_0^0 e^{2\hat g}dr \int_0^{\pi}d \theta \sin \theta =\frac 14\int^{r_{\Sigma}}_0 dr \int_0^{\pi}d \theta \sin \theta  \left[r \hat g^{\prime}+{\hat g}_{,\theta \theta} +r^2\hat g^{\prime \prime}  \right] .
\label{Masazeroazero}
\end{equation}
 We can evaluate the above expression to obtain
\begin{eqnarray}
M&=&\frac 14\int^{r_{\Sigma}}_0 dr \int_0^{\pi}d \theta \sin \theta  \left[\frac{\partial}{\partial r}(r^2 \hat g^{\prime})-r \hat g^{\prime}+{\hat g}_{,\theta \theta}   \right]\nonumber \\
&=&\frac 14 \int_0^{\pi}d \theta \sin \theta \left[\left( r^2 \hat g^{\prime}-r \hat g \right)_0^{r_{\Sigma}}+ \int^{r_{\Sigma}}_0 dr (\hat g +{\hat g}_{,\theta \theta} ) \right].
\label{Masazeroazero1}
\end{eqnarray}
From  (\ref{aygsimpleschw}) we can evaluate the above expression to obtain
\begin{equation}
M=\frac 14\int^{r_{\Sigma}}_0 dr \int_0^{\pi}d \theta \sin \theta  \ (\hat g +{\hat g}_{,\theta \theta} ).
\label{Masazeroazero2}
\end{equation}

The matching conditions and boundary conditions that $G$ must satisfy lead to the following form for $\hat g$:
\begin{equation}
\hat g=(r-r_{\Sigma})^2 H(r^n) (1-y^2) J(y),
\label{goodg}
\end{equation}
with $n\geq 3$ for $H$ being any function of $r^n$ and $J(y)$  any arbitrary function of $y\equiv \cos\theta$, and then (\ref{Masazeroazero2}) leads to:
\begin{eqnarray}
&&M=\frac 14\int^{r_{\Sigma}}_0 dr \int_{-1}^{1}d y \ (1-y^2)\left[  \hat g + (1-y^2){\hat g}_{yy}-y\hat g_y\right]\nonumber \\
&=&\frac 14\int^{r_{\Sigma}}_0 dr \ (r-r_{\Sigma})^2 H\int_{-1}^{1}d y \ (1-y^2)\left[ J (-1+3y^2)-5y(1-y^2) J_y +(1-y^2) J_{yy}\right] \nonumber \\
\label{Masazeroazero3}
\end{eqnarray}

Now, in order to get (\ref{Masazeroazero3}) equal to zero we have two possibilities: either we chose $J$ in such a way that the angular integration vanishes, or the radial integration  goes to zero.  Let us evaluate both scenarios.

Let us start by  considering the first scenario. The angular integration of (\ref{Masazeroazero3}) vanishes if $J$ is an arbitrary  polynomial in odd powers of $y$. Alternatively any combination of even integer powers in $y$ could lead to vanishing mass $M$ iff an  appropriated relation between the coefficients is satisfied. Thus for example,  the function $J(y)=Cy^n+Dy^m$ with $n$ and $m$ being positive even integers, turns (\ref{Masazeroazero2}) to zero,   whenever $C$ and $D$ fulfill the following relation
\begin{equation}
C=-D \frac{(2m-1)(n+5)(n+3)}{(2n-1)(m+5)(m+3)}.
\end{equation}

In the second scenario  we can chose $H=A r^m+B r^k$ with arbitrary constants $A$ and $B$, and $m,k$ being integers equal  to, or greater than $3$. Assuming  $m<k$  we have
\begin{equation}
\int^{r_{\Sigma}}_0 dr (r-r_{\Sigma})^2 H=r_{\Sigma}^{m+3}\int_0^1 ds (s-1)^2(A s^m+Br_{\Sigma}^{k-m}s^k).
\label{radial int}
\end{equation}
The above integral vanishes if the  following condition is satisfied
\begin{equation}
A=-r_{\Sigma}^{k-m} B \frac{k!}{m!}\frac{(m+3)!}{(k+3)!}.
\label{AB}
\end{equation} 

Therefore we can conclude that the function $\hat g$ in the form (\ref{goodg}) with $J$ being a polynomial in odd powers of $y$ and/or $H$ being a lineal combination $H=A r^m+B r^k$ where $A$ and $B$ are related by (\ref{AB}), both lead to an interior  solution describing a stellar object with vanishing total mass $M$ (a ghost star).

In order to specify further our solution  we shall chose a particular simple form for $\hat g$ 
\begin{equation}
\hat g=r_{\Sigma}^{n+2}(s-1)^2 s^n (1-y^2) \kappa y \ , \ n\geq 3,
\label{goodgexample}
\end{equation}
$\kappa$ being an arbitrary parameter, from where we may easily calculate  pressures $\hat p_{xx}$ and $\hat p_{zz}$ to obtain
\begin{equation}
\hat p_{xx}=- \kappa r_{\Sigma}^n s^{n-2} (s-1) y \left[ (s-1)(1+n-(n+3)y^2)+2s(1-y^2)\right],
\label{pxx}
\end{equation}
\begin{eqnarray}
\hat p_{zz}=\kappa r_{\Sigma}^n s^{n-2} y \left\{-2s^2(1-y^2)+\right. \\ \left. \nonumber
  + (s-1)\left[(5-n^2-4n)s+n^2-2n-7+y^2(s(n^2+4n-7)-n^2+9)\right] \right\}.
  \label{pzz}
\end{eqnarray}

Using the above expressions in (\ref{presionPpzero})   we may write 
\begin{equation}
T_0^0=T_3^3=-\frac{e^{-2\hat g}}{8 \pi} \hat p_{zz}\quad  ,\quad  T_1^1=-T_2^2=-\frac{e^{-2\hat g}}{8 \pi} \hat p_{xx}.
\label{tmunu}
\end{equation}

Figure 1	depicts the behavior of  $T_0^0$ and $T_1^1$ in terms of the radial parameter $s$ in the case $n=3$, for different values of the angular variable $y$.

\begin{figure}[h]
	$$
	\begin{array}{cc}
	\includegraphics[scale=0.35]{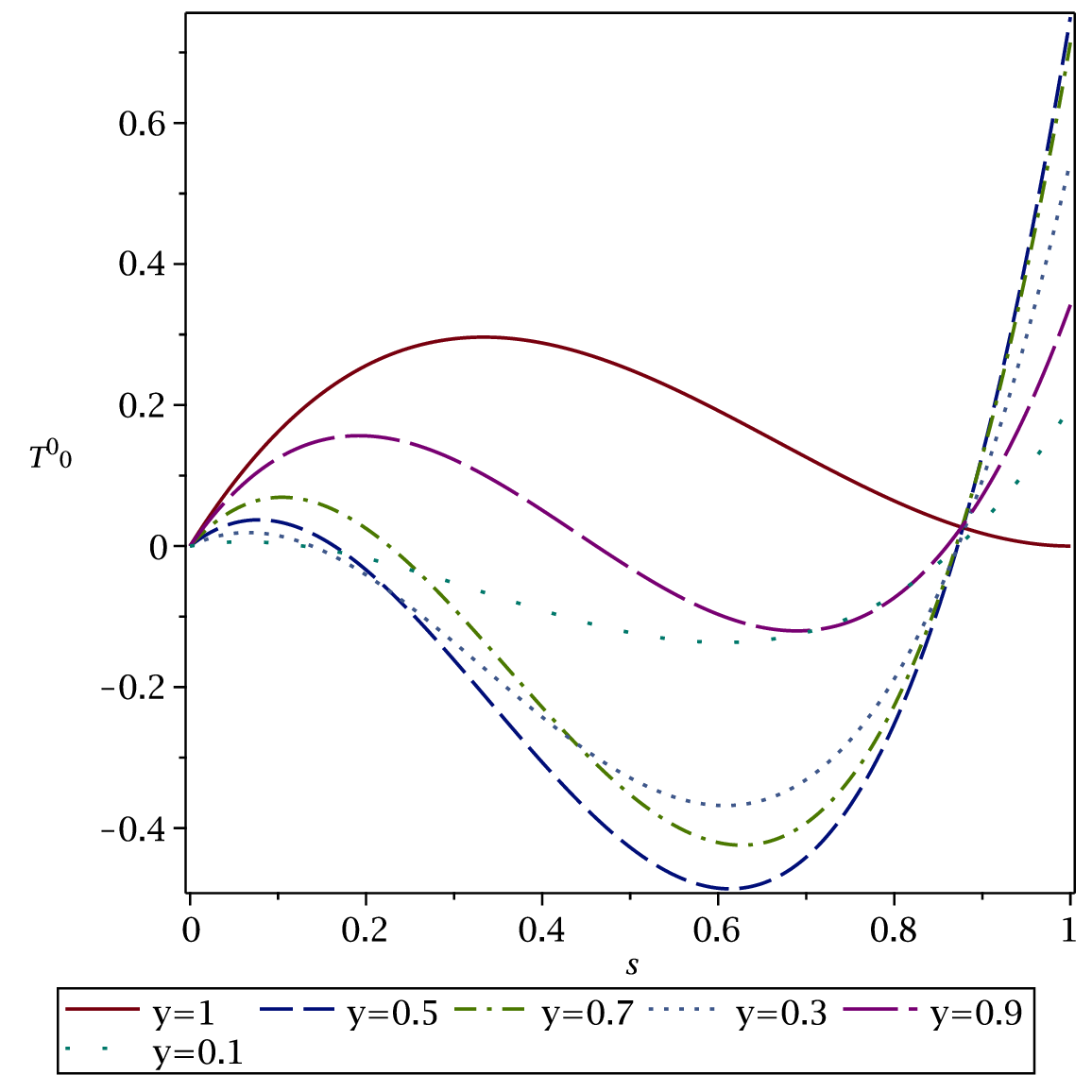}& \includegraphics[scale=0.35]{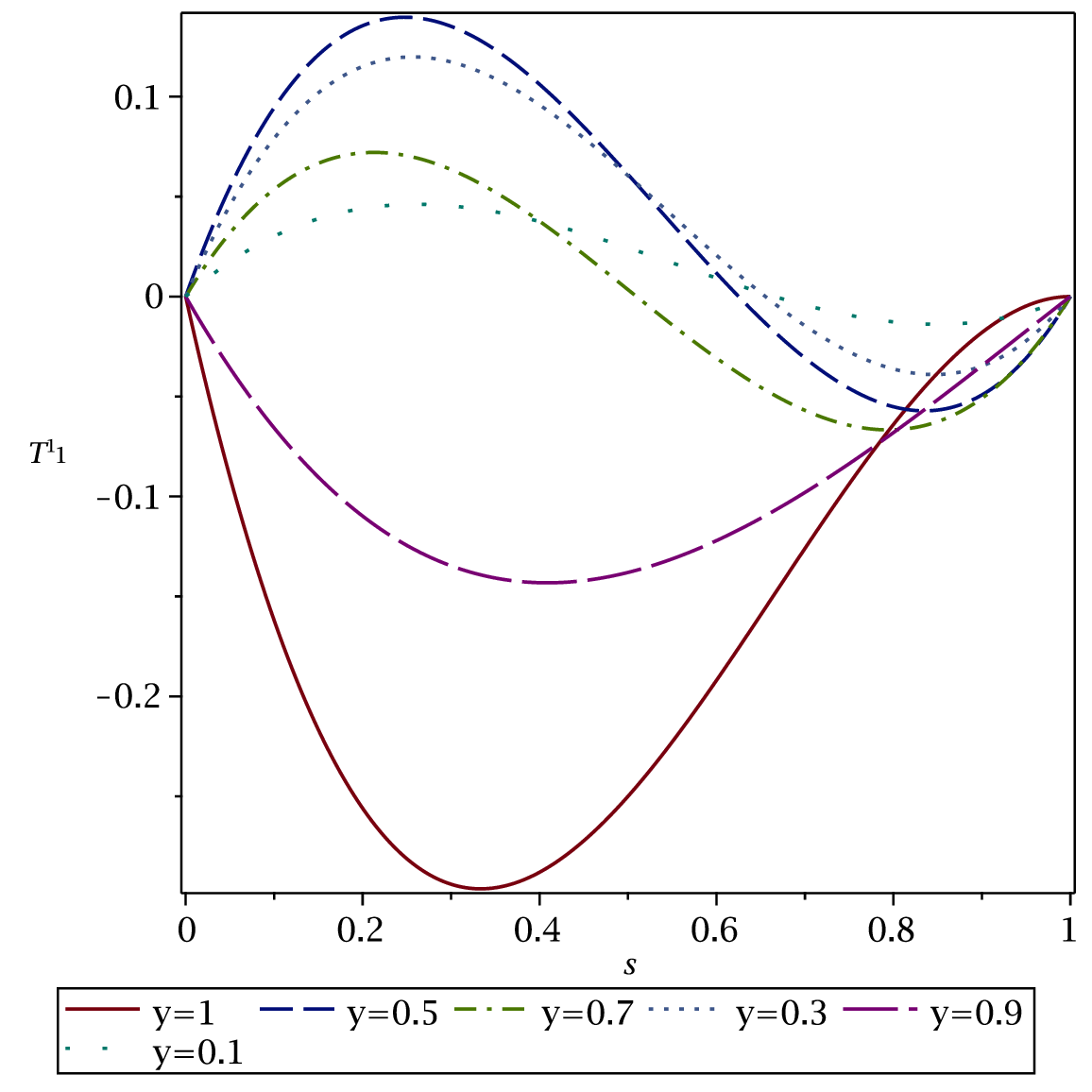} \nonumber\\
	(a) & (b)  \nonumber 
	\end{array}
	$$
	{\caption{\label{modelos} {\it Components of the energy-momentum  (a)  $T_0^0$,   (b)  $T_1^1$ rescaled by a factor $\kappa \frac{r_{\Sigma}^3}{8\pi}e^{-2 \kappa  r_{\Sigma}^5}$,  are represented as functions  of radial variable $s$ and for different values of the angular variable $y$: $y=0.1, 0.3, 0.5, 0.7, 0.9, 1$.} as indicated in the figure} }
\end{figure}

\subsection{Characterization of the geometry of the source }

With the purpose of providing some information about the ``shape'' of our  source, we shall calculate the proper length  $l_z$ of the object  along the axis  $z$, and the proper equatorial radius $l_{\rho}$, given by
\begin{equation}
l_z\equiv\int_0^{r_{\Sigma}}e^{\hat g(y=1)-\hat a(y=1)} dz \ , \  l_{\rho}\equiv\int_0^{r_{\Sigma}}e^{\hat g(y=0)-\hat a(y=0)} d\rho,
\end{equation}
where ${\rho,z}$ are  cylindrical coordinates.

The expressions above for  the proper lengths   allow us to visualize the flattening of the source with respect to the spherical case, and also to relate it with the  parameters $\kappa$ of the source and the free parameter $n$ of the interior metric. 

Let us first notice  that both proper lengths are equal to $r_{\Sigma}$, as it happens in the spherically symmetric  case ($\hat g=\hat a=0$), since for our model (\ref{goodgexample}) the function $\hat g$ vanishes at the axis ($y=\pm 1$) and the equator ($y=0$), whereas   along other angular directions the length is given by 	\begin{equation}
	l(y)\equiv r_{\Sigma}\int_0^{1} ds \ e^{\hat g(y)}=r_{\Sigma}\equiv\int_0^{1} ds \  e^{r_{\Sigma}^{n+2}(s-1)^2 s^n (1-y^2) \kappa y }
	\label{long}
	\end{equation}
The localization of the maximum and minimum values of the above length don't depend on the parameters of the model and they are reached at $y=\pm \frac{1}{\sqrt{3}}$ respectively, i.e. $\theta=\pm \arccos(\frac{1}{\sqrt{3}})$ and $\theta=\pm[\frac{\Pi}{2}+\arccos(\frac{1}{\sqrt{3}})]$, whereas the length of the source for different angular directions varies with the parameters $n$ and $\kappa$ as can be seen in the figure 2. Let us observe that we have introduced a new parameter $\delta\equiv \kappa r_{\Sigma}^{n+2}$ to perform the graphics.

Another feature which deserves to be mentioned is that the maximum (minimum) length increases (decreases) with  the parameter $\delta$. However, the behavior of both lengths with respect to the parameter $n$  is just the opposite  i.e. maximum lengths decrease with increasing $n$ while minimum lengths increase with higher values of the parameter $n$. Figures 3 and 4 show these behaviors by depicting $l(\theta)$ for different values of $n$ and $\delta$.

\begin{figure}[h]
	$$
	\begin{array}{cc}
	\includegraphics[scale=0.35]{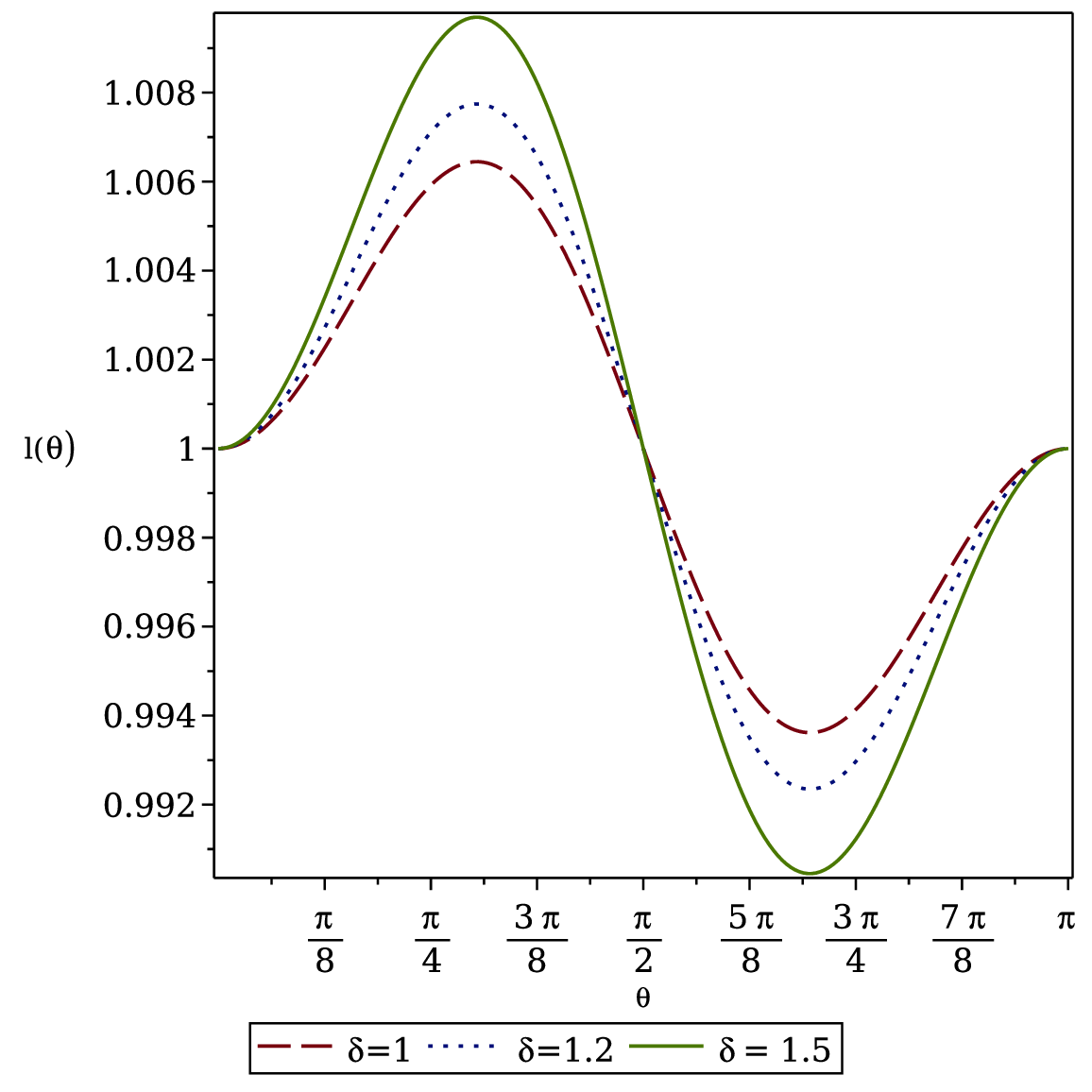}& \includegraphics[scale=0.35]{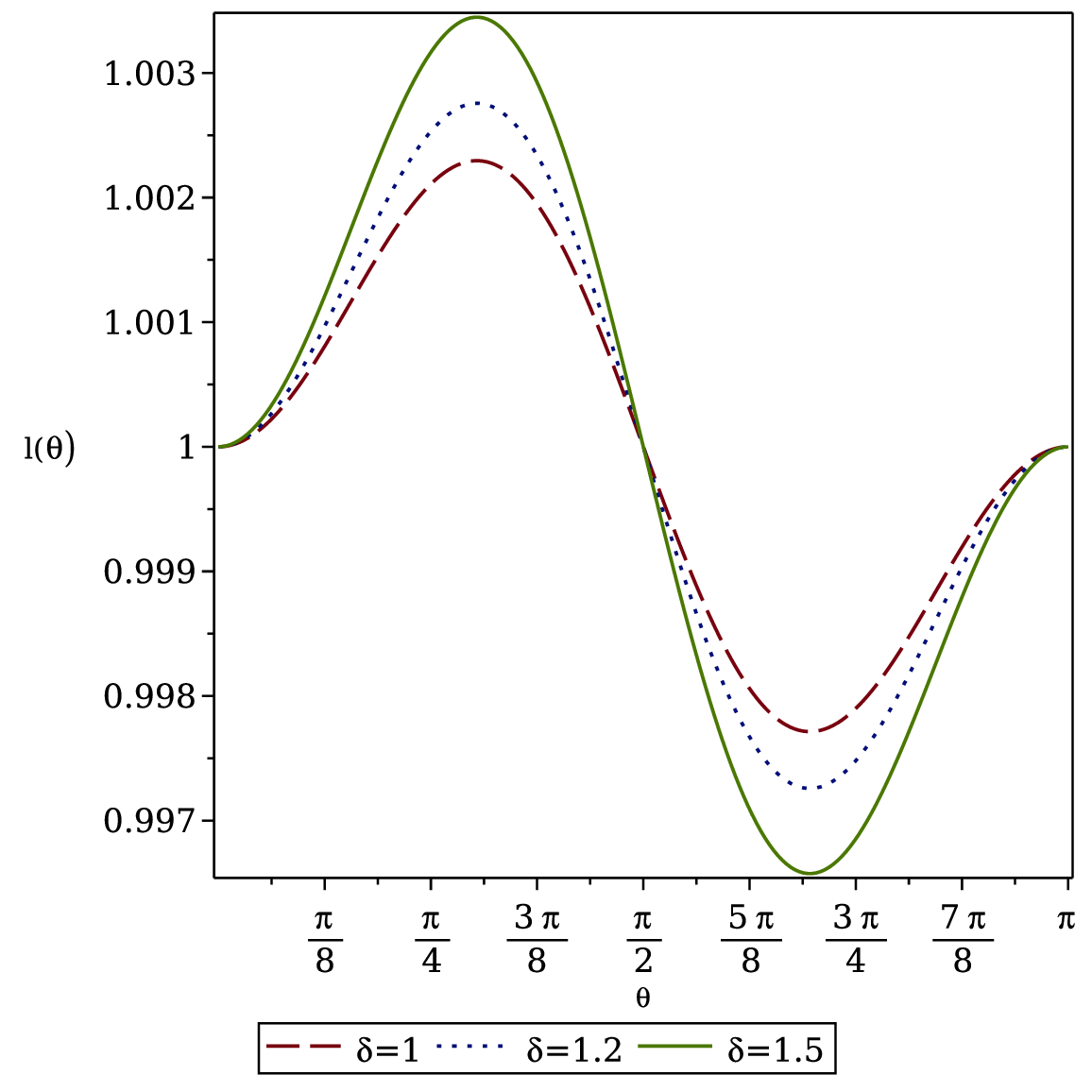}  \nonumber\\
	(a) & (b) \nonumber
	\end{array}
	$$
	\caption{\label{curvas} {\it $l(\theta)$, scaled by a factor $1/r_{\Sigma}$, for different values of $\delta$:  (a) with $n=3$, $\delta=1, 1.2, 1.5$ , and  (b) with $n=5$, $\delta=1, 1.2, 1.5$. }}
\end{figure}

\begin{figure}[h]
	$$
	\begin{array}{cc}
	\includegraphics[scale=0.35]{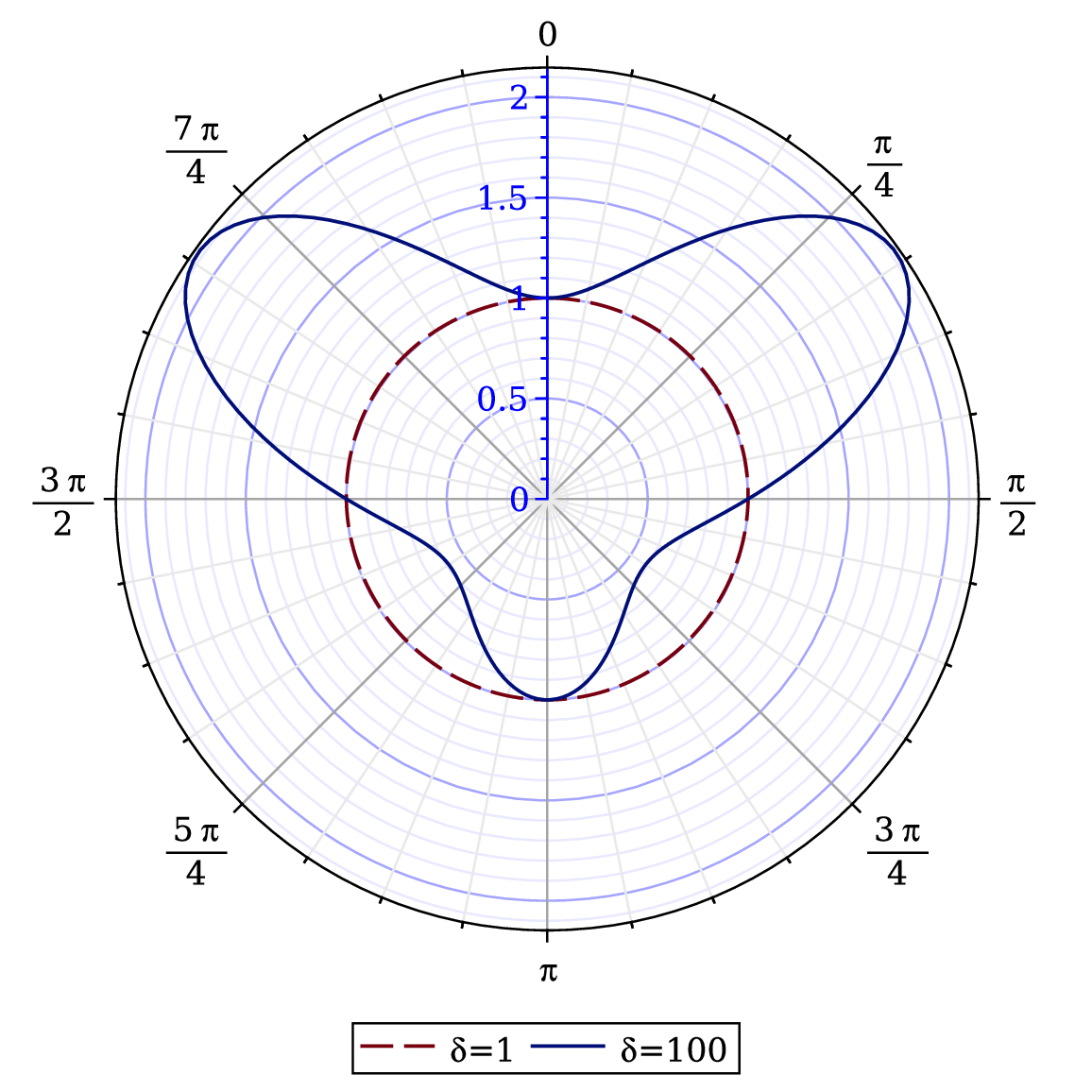}& \includegraphics[scale=0.35]{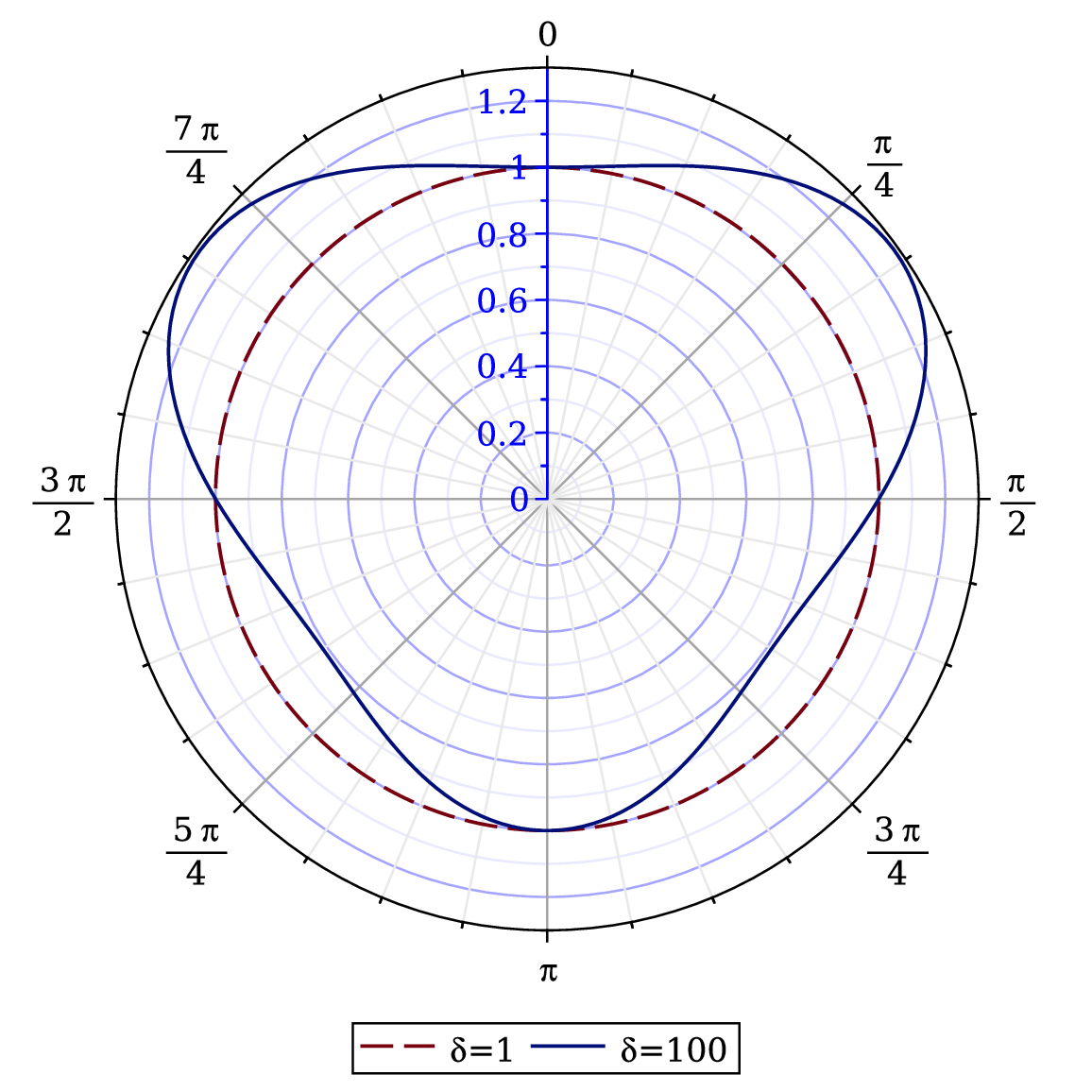}  \nonumber\\
	(a) & (b) \nonumber
	\end{array}
	$$
	\caption{\label{longitudes} The shape of the  source is depicted  by the polar graphic of {\it$l(\theta)/r_{\Sigma}$  for different values of $n$ and $\delta$:  (a) with $n=3$, $\delta=1, 100$, and  (b) with $n=5$, $\delta=1, 100$. Dashed line corresponds to a value of $\delta=1$ whereas the solid line is for $\delta=100$. These values are taken in order to exhibit  the behavior of the maximum and minimum lengths. Dashed line is not a circle as can be observed clearly in figure 2.}}
\end{figure}

\begin{figure}[h]
	\includegraphics[scale=0.35]{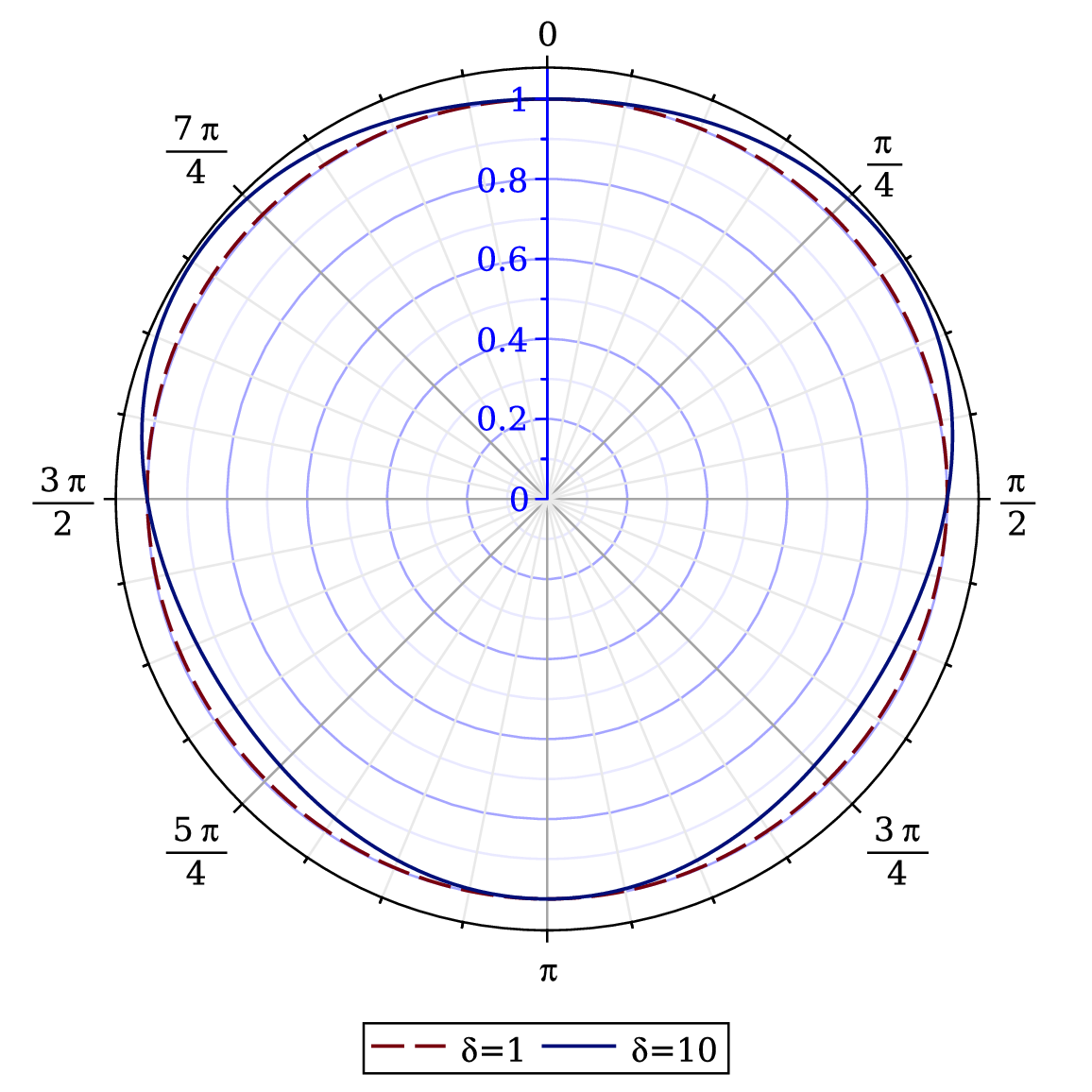}
	\caption{\label{n31a10} The shape of the  source is depicted by the polar graphic of {\it$l(\theta)/r_{\Sigma}$  for different values of $\delta$ with $n=3$, Dashed line corresponds to a value of $\delta=1$ whereas the solid line is for $\delta=10$. Notice  that dashed line is not a circle as can  be observed clearly in figure 2.}}
\end{figure}

\subsection{The complexity of the source}

In recent papers \cite{com1,com2} a new definition of complexity for self--gravitating 	fluids has been proposed, which has been proved to be particularly  suitable for measuring the degree of ``complexity'' of a given fluid distribution. The proposed definition is based on a set of scalar variables called complexity factors, appearing in the orthogonal splitting of the Riemann tensor. More specifically these scalars determine the trace--free electric part of the Riemann tensor.  In the spherically symmetric case \cite{com1} there is only one such a scalar, whereas in the most general axially  symmetric case \cite{com2} there are three of them. 

The electric part of the Riemann tensor $Y_{\mu \nu}$ is defined by
\begin{equation}
Y_{\mu \nu}=R_{\mu \alpha \nu \beta}V^\alpha V^\beta, 
\label{er}
\end{equation}
where $V^\alpha=(1,0,0,0)$ denotes the four--velocity of the fluid in the comoving frame and $R_{\mu \alpha \nu \beta}$ is the Riemann tensor.
A straightforward calculation shows  that, in our case, all the non-vanishing components of the Riemann tensor correspond to latin-indexes (running from $1$ to $3$). 
These components are 
\begin{eqnarray}
R_{1212}&=&-e^{2\hat g}\left( r^2 \hat g^{\prime\prime}+r\hat g^{\prime}+ \hat g_{\theta\theta} \right),\nonumber\\ 
R_{1313}&=&-\sin\theta\left(-r \sin\theta \ \hat g^{\prime}+ \cos\theta \ \hat g_{\theta} \right),\nonumber\\ 
R_{1323}&=&r\sin\theta\left(r \cos\theta \ \hat g^{\prime}+ \sin\theta \ \hat g_{\theta} \right),\nonumber\\ 
R_{2323}&=&r^2\left(r \cos^2\theta \ \hat g^{\prime}+ \sin\theta\cos\theta \ \hat g_{\theta}-r \hat g^{\prime} \right).\nonumber\\ 
\end{eqnarray}

Accordingly  the electric part of the Riemann tensor $Y_{\mu \nu}$ vanishes, implying the vanishing of all complexity factors.

\section{The relativistic multipole moments}

We shall now  obtain some information about our source from the $RMM$ expressed through the variables describing the source. The general theory for doing that has been developed in \cite{whom}, where explicit expressions of $RMM$ in terms of  integrals over the whole space--time has been obtained. 

It is worth mentioning that before \cite{whom}, the usual methods  for the calculation of  $RMM$ involved only the metric of the space-time outside the source e.g.  the Geroch-Hansen \cite{geroch}, \cite{hansen} and Thorne method \cite{thorne}, or the Fodor- Hoenselaers-Perjes method \cite{FHP}. 

Instead the  method presented  in \cite{whom} to calculate $RMM$, is based on  volume integral expressions involving the global metric.

Taking into account the idea expressed above,  we should expect that in our case  all the  $RMM$ cancel out, since  our source is matched to the Minkowski space-time and we know that the flat space-time has no $RMM$,   
			
We shall see below that that such an expectation is justified.

The definition of multipole moments  introduced in \cite{whom} for axially symmetric space-times, is expressed  by means of the following integral
		\begin{equation}
		I_n= \frac{1}{4\pi} \int_V \left[ H_n \hat{\triangle} \xi -\xi \hat{\triangle}(H_n)\right] \sqrt{\hat g}d^3{\vec x},
		\label{MMdef}
		\end{equation}
		where the volumen of integration must be extended to the whole space, and the following notation is used:
		\begin{equation}
		H_n \equiv \frac{(2n-1)!!}{n!} x^{i_1i_2..i_n} e_{i_1i_2..i_n} \qquad , \qquad \xi \equiv\sqrt{-g_{00}},
		\end{equation}
		where $e^{i_1i_2..i_n}\equiv (e^{i_1}e^{i_2}...e^{i_n})^{TF}$ with  $e^k$ being the unit vector along the positive direction of the symmetry axis, and $TF$ denoting its  trace free  part.  The  Laplacian operator is denoted by $\hat{\triangle}\equiv \frac{1}{\sqrt{\hat g}} \partial_k\left(\sqrt{\hat g} \hat g^{kj}\partial_j \right)$, and $\hat g$ is the determinant of the three-dimensional metric.
		Now, the crucial point here is that the integrand in (\ref{MMdef}) is a divergence (see \cite{whom} for details)	and accordingly, that integral can be evaluated either as a volume integral (\ref{MMdef}), or as a surface  integral:
		\begin{equation}
		I_n= \frac{1}{4\pi} \int_{\partial V} \left[H_n \hat g^{kj} \partial_j \xi-\xi \hat g^{kj} \partial_j H_n\right]  d\sigma_k,
		\label{floworvolu}
		\end{equation}
		$\partial V$ being the boundary  and $d \sigma_k$ its corresponding surface element. 
		It is easy to see that the surface integral  (\ref{floworvolu}) leads to the following flux  evaluated at the infinity surface ${\displaystyle F_n^{\infty}(\psi^{\prime})}$ since the integration is done over all the space, 
		\begin{equation}
		F_n(\psi^{\prime})=\frac 12\int_{-1}^{1} r^2 \psi^{\prime} H_n(r) dy  \ ,
		\label{flujoMn}
		\end{equation}
	and therefore all the $RMM$ vanish since the exterior metric function $\psi$ is zero.

Alternatively we can also proceed  to evaluate the volume integral (\ref{MMdef}) with the global metric.
		The integral (\ref{MMdef}) for the  volume extended from the boundary to the infinity, $I_n^E$, is (see (\cite{whom}) for details)
		\begin{equation}
		I_n^E=\frac 12 \int_{r_{\Sigma}}^{\infty}  r^2 dr \int_{-1}^1 dy \left[ \psi_{,\theta}\frac{ H_{n,\theta}}{r^2}+\psi^{\prime}H_n^{\prime}\right],
		\label{VE}
		\end{equation}
		which vanishes since in our case $\psi=0$.
		
		On the other hand  for the interior volume, $I_n^I$, we have
		\begin{equation}
		I_n^I=\int_{0}^{r_{\Sigma}}  \frac{r^2}{2} dr \int_{-1}^1 dy\left\lbrace H_n\left(\hat a^{\prime \prime} +2 \frac{\hat a^{\prime}}{r}\right)+ \hat a ^{\prime} H_n^{\prime} \right\rbrace.
		\label{Vint}
		\end{equation}
	After integration in the radial coordinate the integral for the interior volume finally reduces to
	\begin{equation}
	I_n^I=\frac 12\int_{-1}^1 dy \ {\psi}^{\prime}_ {\Sigma} H_n(r_{\Sigma}),
	\label{Vint2}
	\end{equation} where in the above calculations   the matching conditions $\hat a(r_{\Sigma})=\hat \psi(r_{\Sigma})$, $\hat a^{\prime}(r_{\Sigma})=\hat \psi^{\prime}(r_{\Sigma})$, as well as the assumed behavior at the center $\hat a_0=\hat a^{\prime}_0=0$, have been taken into account. Again, since we are matching with Minkowski $\psi=0$ implying that $I_n^I$ vanishes.
	
	Hence, we can summarize that the volume integral vanishes whatever interior function $\hat a$ with appropriated behavior is chosen (in our case we have considered $\hat a=0$), in concordance with   (\ref{flujoMn}), leading to the vanishing of all $RMM$.

\section{Conclusions}
We have presented  an axially symmetric  
fluid distribution which matches smoothly to the Minkowski space-time on the boundary surface of the source, i.e. the distribution does not produce any gravitational field outside its boundary surface. We called such kind of objects  ``ghost stars'' in \cite{1n} for reasons explained before. 

The obtained solution has all its complexity factors vanishing and in this sense it could be considered as the simplest possible ghost star. Besides, as expected, all its $RMM$ vanish too.

The obtained solution has no spherically symmetric limit as ghost star. This is so because the approach employed for obtaining it assumes that in the spherically symmetric case, the solution corresponds to a homogeneous (in the energy density) and isotropic pressure fluid. However the condition of homogeneity precludes the formation of  a ghost star since there must be a change of sign in the energy density within the fluid distribution, in order to obtain a vanishing total mass.

We should recall that the very existence of ghost stars relies on the assumption of the existence of regions of the fluid distribution endowed with negative energy-density, as depicted in Figure 1 for our model.  In this respect it should be mentioned that negative energy-density (or negative mass) is a subject extensively considered in the literature (see \cite{bon}-\cite{bor} and references therein). Particularly relevant are those references relating the appearance of negative energy-density with quantum effects.

The relevance of the observational aspects of the ghost stars cannot be overemphasized. Thus,   it is evident that the shadow of such kind of object should differ from the one produced by a self-gravitating star with non-vanishing total mass.  We ignore if the ongoing observations of this kind  \cite{et1}-\cite{et4} are able to do that, but this is an important issue to consider. A research endeavor   pointing in a similar direction has been recently published in  \cite{nc}.

On the other hand, it is worth noticing that ghost stars are a sort of reservoirs of dark mass produced by  the appearance of negative energy-density is some regions of the fluid distribution. It remains to be seen if the general problem of dark matter could be, at least partially, be explained in terms of ghost stars \cite{rep}.

\section*{Acknowledgments}
This work was partially supported by the Grant PID2021-122938NB-I00 funded
by MCIN/AEI/ 10.13039/501100011033 and by ERDF A way of making Europe,
as well as the Consejer\'\i a de Educaci\'on of the Junta de Castilla y Le\'on under the
Research Project Grupo de Excelencia  Ref.:SA097P24 (Fondos Feder y en
l\'\i nea con objetivos RIS3).


\end{document}